\documentclass[prl,aps,twocolumn,superscriptaddress,notitlepage]{revtex4-1}
\usepackage{graphicx}
\usepackage{dcolumn}
\usepackage{amsmath}
\usepackage{enumerate,amsthm,amssymb,color}
\usepackage{dsfont}
\usepackage{bbm}
\usepackage{bm}
\usepackage[french, english]{babel}
\usepackage{synttree}
\usepackage{float}
\usepackage{newfloat}
\DeclareFloatingEnvironment[
    fileext=los,
    listname=List of Schemes,
    name=Protocol,
    placement=H,
    within=section,
]{protocol}
\usepackage{adjustbox}

\usepackage{booktabs}

\newcommand{\ie}{{\em i.e.}}

\newcommand{\ket}[1]{|#1\rangle}
\newcommand{\bra}[1]{\langle#1|}

\DeclareMathOperator{\Id}{\mathds{1}}

\graphicspath{{images/}}

\begin{document}


\title{Experimental investigation of practical unforgeable quantum money}

\author{Mathieu Bozzio}\affiliation{LIP6, CNRS, Universit\'e Pierre et Marie Curie, Sorbonne Universit\'es, 75005 Paris, France}\affiliation{LTCI, T\'el\'ecom ParisTech, Universit\'e Paris-Saclay, 75013 Paris, France}
\author{Adeline Orieux}\affiliation{LIP6, CNRS, Universit\'e Pierre et Marie Curie, Sorbonne Universit\'es, 75005 Paris, France}\affiliation{IRIF, CNRS, Universit\'e Paris Diderot, Sorbonne Paris Cit\'e, 75013 Paris, France}
\author{Luis Trigo Vidarte}\affiliation{LIP6, CNRS, Universit\'e Pierre et Marie Curie, Sorbonne Universit\'es, 75005 Paris, France}\affiliation{LCF, Institut d'Optique Graduate School, CNRS, Universit\'e Paris-Saclay, 91127 Palaiseau, France}
\author{Isabelle Zaquine}\affiliation{LTCI, T\'el\'ecom ParisTech, Universit\'e Paris-Saclay, 75013 Paris, France}
\author{Iordanis Kerenidis}\affiliation{IRIF, CNRS, Universit\'e Paris Diderot, Sorbonne Paris Cit\'e, 75013 Paris, France}\affiliation{Center for Quantum Technologies, National University of Singapore, Singapore}
\author{Eleni Diamanti}\affiliation{LIP6, CNRS, Universit\'e Pierre et Marie Curie, Sorbonne Universit\'es, 75005 Paris, France}

\begin{abstract}
\noindent Wiesner's unforgeable quantum money scheme is widely celebrated as the first quantum information application. Based on the no-cloning property of quantum mechanics, this scheme allows for the creation of credit cards used in authenticated transactions offering security guarantees impossible to achieve by classical means. However, despite its central role in quantum cryptography, its experimental implementation has remained elusive because of the lack of quantum memories and of practical verification techniques. Here, we experimentally implement a quantum money protocol relying on classical verification that rigorously satisfies the security condition for unforgeability. Our system exploits polarization encoding of weak coherent states of light and operates under conditions that ensure compatibility with state-of-the-art quantum memories. We derive working regimes for our system using a security analysis taking into account all practical imperfections. Our results constitute a major step towards a real-world realization of this milestone protocol.


\end{abstract}

\date{\today}

\maketitle

The fundamental property of quantum mechanics at the heart of quantum cryptography is the no-cloning theorem~\cite{WZ:nature82}, which states that it is physically impossible to clone an unknown quantum system, that is, to generate two identical copies of the system starting from a single copy. This property in essence prevents a malicious party from recovering information about the system without disturbing it, something that is always possible in the classical world. In his seminal work~\cite{Wie:acm83}, Wiesner used this property to show that by encoding classical information into conjugate quantum bases it is possible to protect the encoded information from forgery. For instance, classical bits can be encoded using the bases $\{\ket{0},\ket{1}\}$ and $\{\ket{+},\ket{-}\}$, with $\ket{\pm} = (\ket{0} \pm \ket{1})/\sqrt{2}$, where the first qubit state in each basis encodes the bit 0 and the second the bit 1. Then, Heisenberg's uncertainty principle ensures that measuring the encoded qubit in one of the two bases destroys any information about the encoding in the other, while the no-cloning theorem ensures that only a party knowing the basis used for the encoding can unambiguously recover the encoded information upon measurement of the qubit. This idea was subsequently extensively used in many quantum cryptographic schemes~\cite{BS:dcc16}, and in particular in the BB84 quantum key distribution protocol~\cite{BB84}, which has since thrived as one of the most studied and successfully implemented quantum information applications~\cite{SBC:rmp09,DL+:npjQI16}.

The application that Wiesner's original work was interested in was unforgeable quantum money. The goal of a quantum money scheme is to perform an authenticated and efficient transaction between a client, a vendor and a bank via the use of a prepaid credit card (private-key money scheme) or between a client and a vendor via the use of banknotes (public-key money scheme), with maximal security guarantees. More precisely, in a quantum credit card system, a trusted bank uses a secret to preload a certain amount of money on a credit card, namely a device that contains encoded quantum information, and gives it to a client who can make payments to a vendor who has in his possession a credit card reader, namely a device that can access the card and communicate with the bank, which can then verify using the initial secret if the credit card is valid and can be used for the payment. The security guarantee is that the client cannot duplicate the credit card or increase the amount of money associated with it. In quantum banknotes, the difference is that the verification must be done without the help of the bank, hence there is no private key involved in the process.

In the classical world, money schemes are impossible with information-theoretic security and are therefore based on computational assumptions. In the quantum world, schemes for unforgeable quantum credit cards typically involve verification procedures with quantum communication with the bank~\cite{Wie:acm83}. A simpler information-theoretically secure credit card scheme using classical communication during verification was proposed~\cite{Gav:ccc12}, and is based on hidden matching quantum retrieval games~\cite{BJK:stoc04,AKL:pra16}; these typically involve several rounds of communication between the vendor and the bank and the use of specific entangled states, although these requirements have been improved recently~\cite{AA:pra17}. This classical verification scheme has also been further simplified to use again the so-called BB84 states as in the original Wiesner scheme, additionally taking into account realistic conditions~\cite{PY+:pnas12}. As regards to quantum banknotes, it is known that even in the quantum world such schemes cannot base their security solely on the no-cloning theorem but must also use some computational assumptions, such as knot problems or quantum obfuscation~\cite{LA+:ics10,MS:ecc10,AC:stoc12,FG+:itcs12,GK:tqc15,AF:arxiv16}. This computational security is still interesting since in the classical world there can be no notion of mathematical security for banknotes; their security is based only on the fact that it is difficult for a counterfeiter to copy a banknote due to its intricate coloring and hologram design. Recent experimental work has shown how such quantum banknotes can be constructed on-the-fly but also forged~\cite{BC+:npjQI17}. On the other hand, unforgeable quantum credit card schemes have never been implemented until now.

Here we develop and implement an on-the-fly version of a practical unforgeable quantum money scheme for credit card transactions for the first time (independent work implementing a hidden matching version of quantum money has appeared since in~\cite{GAA:arx17}). Our protocol builds upon the work in~\cite{PY+:pnas12} in conjunction with the techniques developed in~\cite{GK:tqc15}, and uses only BB84 states and a single round of classical communication for the verification. Our construction allows us to formulate conditions for correctness and security that can be experimentally tested. In particular, we derive conditions in scenarios of practical interest, corresponding to quantum memories based on single emitters or on atomic ensembles~\cite{QMreview2009,QMreview2015,HE+:jmo16}, and to implementations with weak coherent states.
In all cases, we demonstrate that these conditions are rigorously satisfied experimentally using a practical photonic setup based on polarization encoding of weak coherent states of light, and we analyze operational regimes where unforgeability is guaranteed. Our experiment includes the full procedure of credit card state generation, readout and verification and is compatible with the future use of quantum storage devices, hence paving the way for the realization of quantum money transactions with information-theoretic security impossible to achieve in the classical world.

\medskip

\begin{figure}
\includegraphics[width=8cm]{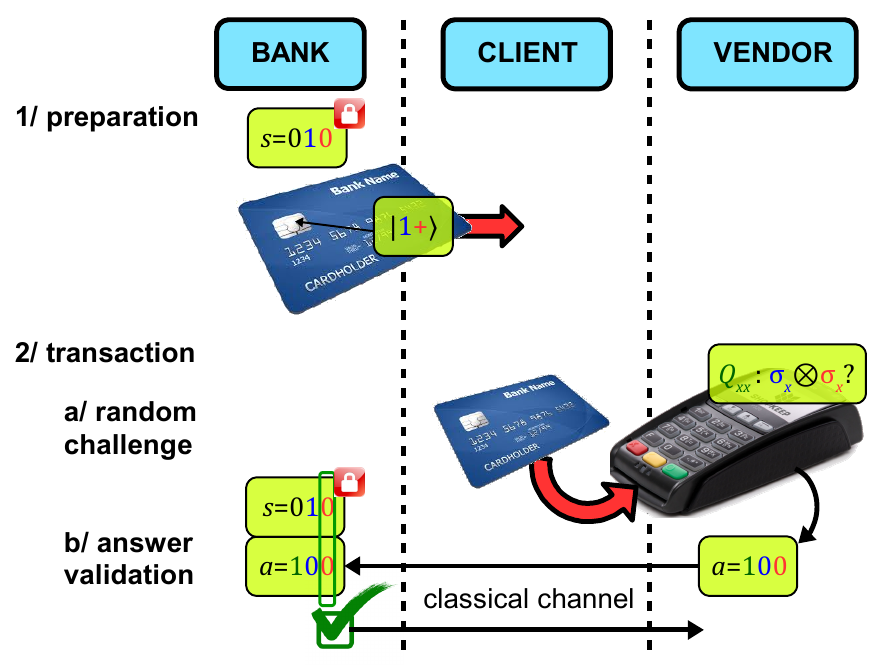}
\caption{\textbf{Practical quantum money protocol.} The sequence of interactions between the credit card holder (client), the bank and the vendor involved in the transaction. In the preparation phase, the bank uses a secret key to prepare the quantum state loaded on the credit card, which is then given to the client. In the transaction phase, the vendor randomly selects one out of two challenge questions, measures the qubits and sends the outcome to the bank, who can then verify the validity of the credit card or detect a forgery attempt.}\label{fig:protocol}
\end{figure}

\noindent {\large \textbf{Results}}\\
\noindent\textbf{The quantum money protocol.} To describe the protocol that we have analyzed, we first define a reduced scheme that guarantees a weak security condition, before extending it to a larger scheme with exponentially good security parameters. The reduced scheme is schematically shown in Fig.~\ref{fig:protocol}.

Let us assume that the bank prepares a prepaid credit card state, which for now consists of a single pair of qubits only, chosen from the following set using a random secret classical string $s$ :
\begin{equation}
S_{\text{pair}}=\{\ket{0+},\ket{0-},\ket{1+},\ket{1-},\ket{+0},\ket{+1},\ket{-0},\ket{-1}\},
\label{eq:Spair}
\end{equation}
where $\ket{0}$, $\ket{1}$ and $\ket{+}$, $\ket{-}$ are the Pauli $\sigma_{z}$ and $\sigma_{x}$ basis eigenstates, respectively. The secret string $s$ consists of three bits, $\{b,c_0,c_1\}$, where $b=0$ indicates that the first qubit is encoded in the $\sigma_{z}$ basis and the second in the $\sigma_{x}$ basis, while $b=1$ indicates that the first one is encoded in the $\sigma_{x}$ basis and the second in the $\sigma_{z}$ basis. $c_0,c_1$ are the two encoded bits, with $0$ corresponding to the states $\ket{0},\ket{+}$ and $1$ corresponding to the states $\ket{1},\ket{-}$.

When a transaction is to be made, the following interactions occur: first, the client hands the credit card to the vendor, who chooses at random one out of two \emph{challenge questions} and accesses the credit card, \emph{i.e.}, performs a measurement on the stored qubits in order to get an answer for the selected challenge. As a second step, the vendor sends the classical bits corresponding to the chosen challenge, along with the answer obtained upon measurement, to the bank. Finally, the bank verifies the authenticity of the credit card using the secret string $s$ and responds with a yes or no. If the answer of the bank is yes then the transaction may occur, otherwise the card is rejected and declared as a counterfeit.

The two challenge questions, $Q_{zz}$ and $Q_{xx}$, read:

\medskip
\noindent $Q_{zz}$ = \textit{Guess the two bits $c_0,c_1$, such that the guess corresponding to the qubit prepared in the $\sigma_{z}$ basis is correct.}

\medskip
\noindent $Q_{xx}$ = \textit{Guess the two bits $c_0,c_1$, such that the guess corresponding to the qubit prepared in the $\sigma_{x}$ basis is correct.}
\medskip

We may also define the conjunction of both challenges:

\medskip
\noindent $Q_{\epsilon}$ = \textit{Guess the two bits $c_0,c_1$.}
\medskip

The main idea here is that a valid credit card can always be verified in the ideal case: the vendor performs a measurement in the $\sigma_{z} \otimes \sigma_{z}$ basis in order to verify the $Q_{zz}$ challenge, and in the $\sigma_{x} \otimes \sigma_{x}$ basis in order to verify the $Q_{xx}$ challenge. In general, we denote by $c$ the probability of successfully answering $Q_{zz}$ or $Q_{xx}$, and we call this the \emph{correctness parameter}. In the ideal case, $c$ is equal to 1 for the above challenges: measuring both qubits in the $\sigma_{z}$ basis always answers the $Q_{zz}$ challenge correctly, and similarly for $Q_{xx}$. In a realistic implementation, however, $c$ might not be equal to 1 due to system imperfections. It might also take different values, $c_{zz}$ and $c_{xx}$, for $Q_{zz}$ and $Q_{xx}$, respectively. In this case, $c$ can be calculated as the average of $c_{zz}$ and $c_{xx}$, as we describe later.

Let us now see what happens if a dishonest client tries to duplicate the credit card. Since the 2-qubit state is unknown, there is no way that the two copies of the credit card can pass both challenges with high probability. The upper bound on this cheating probability has been derived in ~\cite{Gav:ccc12,PY+:pnas12} 
and shown to be $3/4$. A crucial property of the game is that if one plays in parallel $t$ such games, then one can upper bound the probability a dishonest client can answer the challenge $Q_{\epsilon}$ for all $t$ games as $(3/4)^t$ ~\cite{Gav:ccc12}. In other words, performing a general attack on the composite Hilbert space of all qubit pairs in the card cannot yield a higher cheating probability than performing an optimal attack on each individual pair. In general, we denote by $\epsilon$ the probability of successfully answering the conjunction challenge $Q_{\epsilon}$, \emph{i.e.}, the probability of successfully cheating, and we call this the \emph{security parameter} materializing the non-clonability of the quantum state. We refer to such a challenge game as a ($c$,$\epsilon$)-game $G$.

Starting from a $(c,\epsilon)$-game $G$ that consists of a single pair of qubits and for which the following relation between the correctness and the security parameter holds,
\begin{equation}
c>\frac{\epsilon +1}{2},
\label{eq:cepsilon}
\end{equation}
we can construct a different game $G'$ that consists of $n$ such pairs of qubits, where we denote by $\ket{\$_{s}}$ the $2n$-qubit state chosen by a randomly generated secret classical string $s$ of length $3n$.
Here again, the vendor chooses at random one out of two challenge questions and performs a measurement on all $2n$ qubits accordingly (either $\sigma_{z} \otimes \sigma_{z}$ on all $n$ pairs or $\sigma_{x} \otimes \sigma_{x}$ on all $n$ pairs).
If we now define the two challenges as: (i) answering the challenge $Q_{zz}$ correctly for at least a fraction $c-\delta$ of the $n$ pairs, and (ii) answering the challenge $Q_{xx}$ for at least a fraction $c-\delta$ of the $n$ pairs, where we define (see Methods)
\begin{equation}
\delta =\frac{2c-\epsilon-1}{3}>0,
\label{eq:delta}
\end{equation}
then we can use the analysis in~\cite{GK:tqc15}, based on a Chernoff bound argument~\cite{MU}, and have that the correctness $c'$ and security $\epsilon'$ of the game $G'$ satisfy
\begin{equation}
c'=1-e^{-\frac{cn}{2}\delta^{2}} \mbox{ and } \epsilon'=e^{-\frac{\epsilon n}{3}\delta^{2}}.
\label{eq:cepsilonprime}
\end{equation}
This means that, for the game $G'$, the correctness parameter $c'$ is exponentially close to 1 and the security parameter $\epsilon'$ exponentially close to zero.

Since in our initial game $G$ we have $\epsilon = 3/4$, then we can see from Eq.~(\ref{eq:cepsilon}) that in any secure implementation of the game $G$ we need to achieve $c > 7/8 = 0.875$. The more $c$ exceeds this bound, the better security (\emph{i.e.}, the lower $\epsilon'$) we will get for a game $G'$ of size $n$.

The above description provides a game with exponentially good security parameters. By including quantum states that correspond to many such games in the same credit card as well as a unique classical serial number, one can use theorems from~\cite{Gav:ccc12,AC:stoc12,GK:tqc15} to extend the above scheme into a quantum prepaid credit card scheme, where the quantum credit card may be re-used multiple times and a dishonest client cannot create a copy of the credit card even if he has in his possession multiple credit cards.
Hence, satisfying Eq.~(\ref{eq:cepsilon}) experimentally is enough to implement a full quantum money scheme with information-theoretic security given by correctness and security parameters from Eq.~(\ref{eq:cepsilonprime}).\\


\noindent\textbf{Security analysis for weak coherent states.} In our discussion up till now, we have assumed that the bank creates single-qubit states and stores them in the credit card. In practice, this assumption would be compatible with an implementation using either a quantum memory based on single emitters~\cite{Rempe2011,Lukin2012}, which are expected to emit true single photons to be measured by the vendor for verification regardless of the input state used by the bank in the card preparation stage, or a quantum memory based on atomic ensembles~\cite{NV+:natphoton14,Pan2015} when the input state is a true single-photon state. In the following, we shall refer to this case as the ``single-photon state'' protocol. The correctness and security parameters defined above apply to this case.

In other cases of practical interest, however, we would like to use atomic-ensemble quantum memories and also weak coherent states as an input, as is typically the case, for instance, in quantum cryptographic applications~\cite{SBC:rmp09}. In this case, the memory preserves the Poisson photon statistics in the output state and simply introduces attenuation, hence reducing the average photon number per pulse $\mu$ that characterizes such states. The security threshold therefore has to be modified. More specifically, the bound that $c$ must exceed has to be a function of $\mu$. In the following, we shall refer to this case as the ``weak coherent state'' protocol. Our security analysis first considers specific attacks that may take place in an experimental implementation where the phase of each state is not randomized, namely unambiguous state discrimination (USD) attacks. As a second step, we derive a rigorous bound that applies to all attacks in the phase randomized case.

Starting with the non phase-randomized case, USD attacks are possible only for sets of linearly independent states~\cite{C:pla98,CB:pla98,DJL:pra00}, which is the case for the set of states used in our protocol when physically realized with weak coherent state encoding. A dishonest client willing to copy the credit card can perform specific positive operator valued measurements (POVM) to perfectly discriminate and identify a fraction of the states in the card. Successfully identifying one state in a pair allows the successful cloning of the whole pair, since the adversary knows that the other state is prepared in the conjugate basis. Following the analysis in Refs.~\cite{CB:pla98} and~\cite{DJL:pra00} for our set of states, for $\mu \leq 2$, and assuming that no phase randomization is performed on our states, gives a probability for successful USD
\begin{equation}
P_{D} = 2 e^{-\frac{\mu}{2}} \left(\sinh{\frac{\mu}{2}}-\sin{\frac{\mu}{2}}\right).
\end{equation}

By a Chernoff bound argument~\cite{MU}, we then have that for $n$ pulses (in the $2n$-qubit sequence) that are created according to the Poisson distribution with a mean photon number $\mu$, with very high probability the number of pulses among these $n$ pulses for which the USD is successful is very close to its expectation. In other words, if $L_1,...L_n$ are random variables that take the value 1 if the pulse leads to a successful USD, then we have for the sum $L=\sum_i L_i$ that
\begin{equation}\label{eq:probL}
\Pr[L \geq (1+\eta) P_{D}] \leq e^{- \frac{P_{D}n}{3} \eta^2},
\end{equation}
where $\eta > 0$ is a parameter accounting for finite number statistics that can be optimized as discussed further on.

We can now define a new parameter $\delta$ as
\begin{equation}
\delta =\frac{2c-\epsilon-(1+\eta)P_{D}-1}{3}>0,
\label{eq:delta2}
\end{equation}
and restate the condition of Eq.~(\ref{eq:cepsilon}) as
\begin{equation}
c>\frac{\epsilon + (1+\eta)P_{D} +1}{2},
\label{eq:cepsilon2}
\end{equation}
leading to the following correctness and security parameters that take into account possible USD attacks:
\begin{equation}
c'=1-e^{-\frac{cn}{2}\delta^{2}} \mbox{ and } \epsilon' \leq e^{-\frac{\epsilon n}{3}\delta^{2}} + e^{-\frac{P_{D}n}{3}\eta^2}.
\label{eq:cepsilonprime2}
\end{equation}
Note that the second term in the expression of $\epsilon'$ comes from the fact that, in case $L$ is bigger than its expectation, then the dishonest client can perfectly cheat on a larger number of pairs.

A dishonest client may further boost his cheating probability by exploiting the losses present in a realistic implementation. In the presence of losses, the client having used USD to copy the card may indeed replace the states that he has successfully identified with states containing a higher average photon number $\mu$ at the input of the card reader, in order to increase the probability of detection by the vendor, and replace the ones that he failed to identify with vacuum states. Such a strategy will not be detected when a state is measured in the correct basis but it will induce an increase in the number of total clicks on the detectors registered by the card reader when a state is measured in the conjugate basis. As was shown in~\cite{CB:pla98}, in order for this cheating strategy to be detected and thus for the protocol to be secure, the total efficiency must fulfill the condition
\begin{equation}
\eta_{\text{qm}}\eta_{\text{det}}>\frac{-\textrm{ln}(1-P_{D})}{\mu},
\label{eq:pertes}
\end{equation}
where $\eta_{\text{qm}}$ is the retrieval efficiency of the quantum memory in the original valid card and $\eta_{\text{det}}$ is the detection efficiency of the card reader.

Thus, for the ``weak coherent state'' protocol, as long as it is possible to satisfy Eqs.~(\ref{eq:cepsilon2}) and (\ref{eq:pertes}) experimentally, then a quantum credit card scheme secure against USD attacks, characterized by correctness and security parameters given in Eq.~(\ref{eq:cepsilonprime2}), can be implemented.\\


\begin{figure*}
\centering
\includegraphics[width=\textwidth]{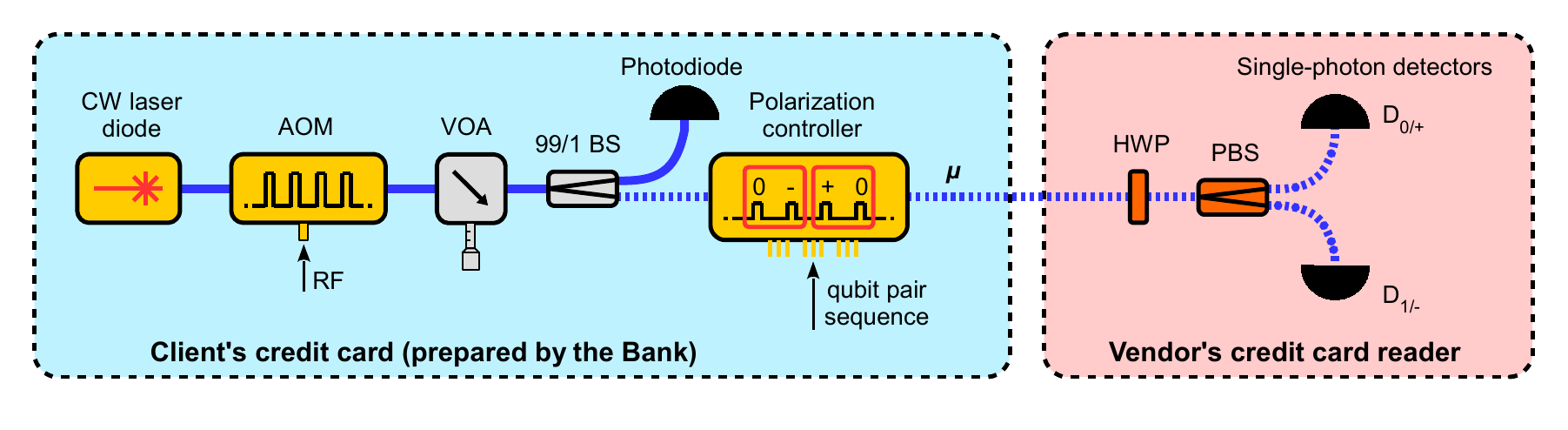}
\caption{\textbf{Experimental setup of the quantum money system.} The credit card state preparation is performed using pulses carved from light emitted by a telecommunication wavelength laser diode using an acousto-optic modulator (AOM). A multi-stage polarization controller (EOSPACE) is then used to select the polarization states according to the protocol by applying suitable voltages. The average photon number of pulse $\mu$ is set by a variable optical attenuator (VOA) and is calibrated with a 99/1 beam splitter (BS) and a photodiode. The credit card reader is materialized by a standard polarization analysis setup including a half-wave plate (HWP), a polarization beam splitter (PBS) and two InGaAs single-photon avalanche photodiodes (ID201). The entire setup is synchronized using a multi-channel delay generator and is controlled by software incorporating the random state generation and data acquisition and processing.}
\label{fig:expsetup}
\end{figure*}

In the case where phase randomization is performed, then each weak coherent state appears to the adversary as a Poisson distributed mixture of Fock states~\cite{LP:calt05}:
\begin{equation}
    \frac{1}{2\pi}\int_{0}^{2\pi} \ket{\sqrt{\mu} e^{i\phi}}\bra{\sqrt{\mu} e^{i\phi}}d\phi = e^{-\mu}\sum_{n=0}^{\infty}\frac{\mu^n}{n!}\ket{n}\bra{n}
\end{equation}

This allows us to derive a general security bound considering three distinct cases. If the state is a vacuum state, then there is no information content. If the state is a single-photon state, then the previous security analysis may be used against general attacks. If the state contains 2 photons or more, then we assume that the adversary may perfectly cheat. This allows us to derive a more pessimistic yet rigorous security threshold,
\begin{equation}
c>\frac{\epsilon + (1+\eta)\lambda +1}{2},
\label{eq:cepsilonlambda}
\end{equation}
where $\lambda=\frac{1-(1+\mu) e^{-\mu}}{1-e^{-\mu}}$, and
\begin{equation}
c'=1-e^{-\frac{cn}{2}\delta^{2}} \mbox{ and } \epsilon' \leq e^{-\frac{\epsilon n}{3}\delta^{2}} + e^{-\frac{\lambda n}{3}\eta^2}.
\label{eq:cepsilonprimelambda}
\end{equation}

As long as it is possible to satisfy Eq.~(\ref{eq:cepsilonlambda}) experimentally, then a quantum credit card scheme with information-theoretic security against all attacks performed on uniformly phase randomized weak coherent states, characterized by correctness and security parameters given in Eq.~(\ref{eq:cepsilonprimelambda}), can be implemented.\\

\noindent\textbf{Experimental principle and setup.} In order to test in practice the security conditions from Eqs.~(\ref{eq:cepsilon}), (\ref{eq:cepsilon2}) and (\ref{eq:pertes}) and identify suitable operation regimes, we have implemented an on-the-fly version of our quantum money protocol in which the qubit pairs of the credit card are sent directly to the vendor's card reader, without intermediate storage in a quantum memory. In our experiment we have not performed phase randomization (this can be implemented subsequently in the same manner as ~\cite{ZQL:apl07}), hence the security bound of Eq.~(\ref{eq:cepsilonlambda}) has not been explicitly considered. The verification test consists in measuring the correctness parameters $c_{zz}$ for the challenge question $Q_{zz}$ and $c_{xx}$ for $Q_{xx}$ on blocks of $n$ qubit pairs, each of which is randomly chosen from the set of Eq.~(\ref{eq:Spair}). This is done by measuring each block in the $\sigma_{z}\otimes\sigma_{z}$ basis or in the $\sigma_{x}\otimes\sigma_{x}$ basis, respectively. As noted earlier, the correctness parameter is calculated as $c = (c_{zz} + c_{xx})/2$, and must exceed the thresholds from Eqs.~(\ref{eq:cepsilon}) and (\ref{eq:cepsilon2}) in order for the credit card to be validated by the bank. This does not compromise the security of the implementation as it is always possible to symmetrize the data by relabeling the bases such that in practice the two parameters become effectively the same. Once $c$ has been measured, the correctness and security of the full protocol can be estimated for the different scenarios from Eqs.~(\ref{eq:cepsilonprime}) and (\ref{eq:cepsilonprime2}).

The experimental setup is shown in Fig.~\ref{fig:expsetup}. The qubit pairs of the credit card state are encoded in the polarization of weak coherent states of light produced with standard optical communication components. The light emitted at $1564$~nm by a continuous-wave laser diode is first modulated using an acousto-optic modulator to produce pulses with a duration of $20\,\mu$s and a repetition rate of $20$~kHz. A variable optical attenuator is used to reduce the intensity of the pulses and set the average photon number per pulse $\mu$. Then, the light pulses go through a multi-stage polarization controller consisting of an electro-optic modulator, which sets the polarization of each pulse to horizontal, vertical, diagonal or antidiagonal, according to a suitable combination of applied voltages. These polarization states correspond to the qubit states $\ket{0}$, $\ket{1}$, $\ket{+}$ and $\ket{-}$, respectively. The voltage sequences applied to the controller are generated such that two successive pulses form a pair whose polarization state is randomly chosen from the set $S_{\text{pair}}$, as required by our protocol.

The value of $\mu$ is fixed for each experiment at the output of the polarization controller, as described in more detail in the Methods. Finally, the pulses are directed to the credit card reader, where the polarization state of each pulse is measured in either the $\sigma_z$ or the $\sigma_x$ basis by the combination of a half-wave plate, set at an angle $0^\circ$ or $22.5^\circ$ with respect to the horizontal direction, respectively, and a polarization beam splitter whose outputs are directed to two single-photon detectors labeled $D_{0/+}$ and $D_{1/-}$. At the end of the experiment, the data sets corresponding to the credit card state generated by the bank, the bases selected by the vendor, and the measurement outcomes obtained for the different challenge questions are analyzed to assess the security of our implementation.


\begin{figure*}
\centering
\includegraphics[width=0.8\textwidth]{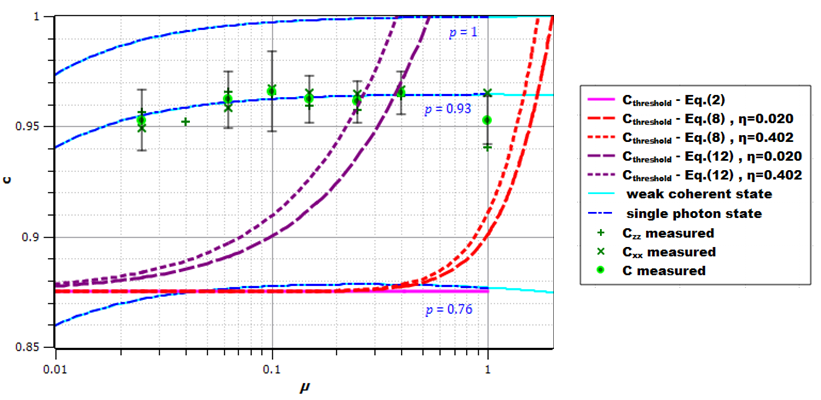}
\caption{\textbf{Experimental results for different values of $\bm \mu$.} Measured $c_{zz}$, $c_{xx}$, and $c$ values (green symbols) are plotted as a function of the average photon number per pulse $\mu$. Each measured block consists of a number of post-selected pairs ranging from $1.3\times 10^{5}$ for $\mu=0.025$ to $2.6\times 10^{5}$ for $\mu=0.40$ and $2.0\times 10^{5}$ for $\mu=1$. The red lines correspond to the security threshold for the ``weak coherent state'' protocol encompassing USD attacks for values of $\eta=0.020$ and $\eta=0.402$, while the purple line corresponds to the threshold for general attacks on phase-randomized weak coherent states, for the same values of $\eta$. The full pink line corresponds to the security threshold for the ``single-photon state'' protocol. The cyan curves correspond to theoretical simulations for weak coherent states assuming a dark count probability of $7\times 10^{-5}$, detection efficiency of $25\%$, state purity values of $p=0.76, 0.93, 1$ and post-selection of pulses with at least one detector clicking. The blue curves correspond to the same theoretical simulations, this time for true single-photon states with an emission efficiency $\mu$. The plotted error bars correspond to $5\sigma$ (see Methods).}
\label{fig:result2}
\end{figure*}

\begin{figure}
\centering
\includegraphics[width=0.50\textwidth]{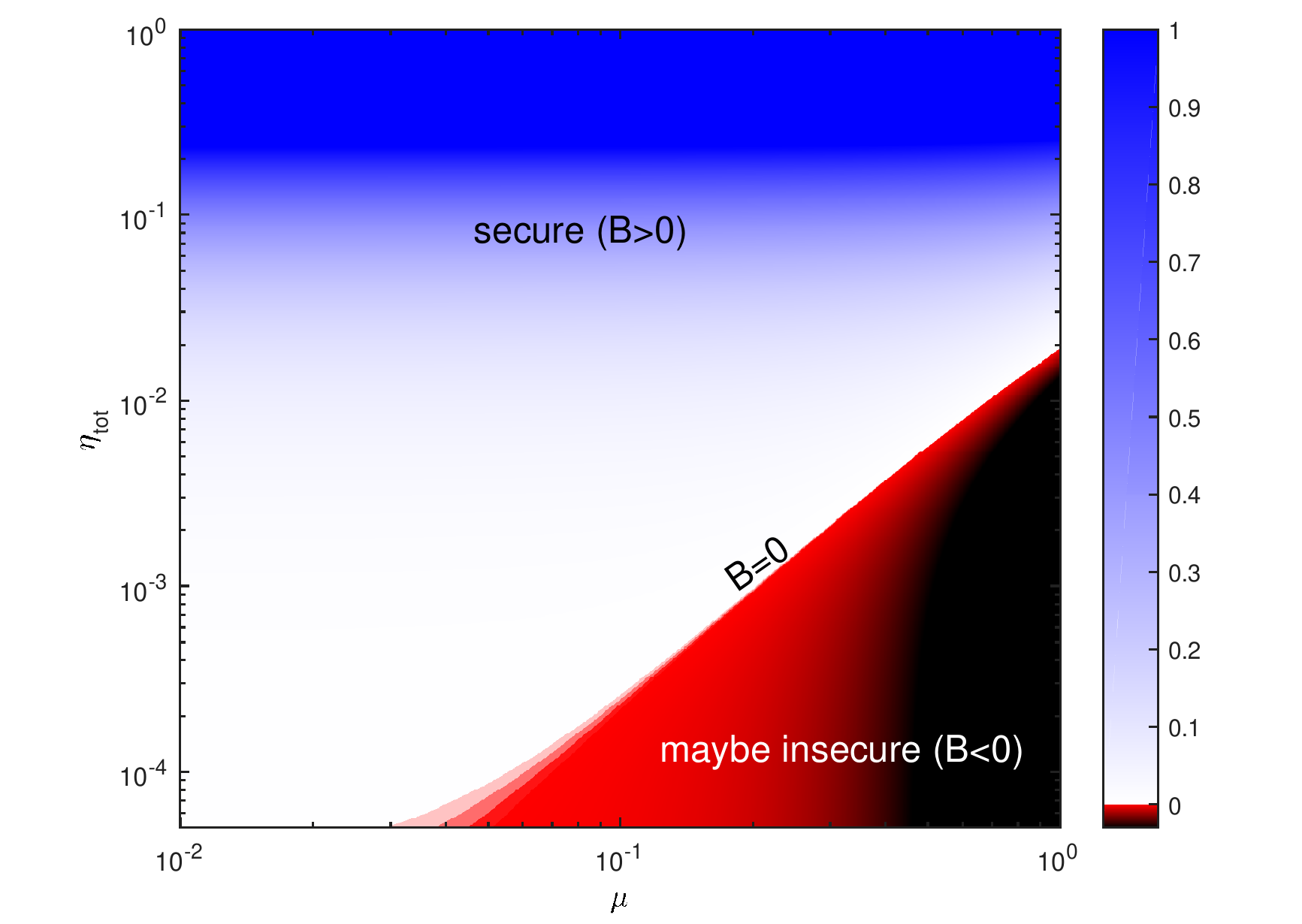}
\caption{\textbf{Security regions for weak coherent states.} $B=\eta_\textrm{tot}+\ln(1-P_D)/\mu$ is plotted as a function of the average number of photons per pulse $\mu$ and the total efficiency $\eta_\textrm{tot}=\eta_{\text{qm}}\eta_{\text{det}}$. The security condition of Eq.~(\ref{eq:pertes}) is fulfilled when $B>0$.}
\label{fig:losses}
\end{figure}

\noindent \textbf{Quantum money results.} The experimental results for the values of $c_{zz}$, $c_{xx}$ and $c$, obtained for weak coherent states with different values of $\mu \leq 1$, are shown in Fig.~\ref{fig:result2} (green symbols). We also display the security thresholds corresponding to Eq.~(\ref{eq:cepsilon}) for the ``single-photon state'' protocol (full pink line) and Eqs.~(\ref{eq:cepsilon2}) and ~(\ref{eq:cepsilonlambda}) for the ``weak coherent state'' protocol, which is plotted for different values of the parameter $\eta$ (dashed red and purple lines respectively). We post-select on events for which at least one of the detectors has clicked (see Methods for more details on the extraction of these parameters).
We have also plotted simulations of the evolution of $c$ with $\mu$ (cyan lines) according to a theoretical model that takes into account Poisson statistics of weak coherent states, dark count probability, finite detection efficiency, state purity and post-selection of pulses where at least one detector clicks (see Methods for more details). The best fit of our data points corresponds to a state purity $p=93\%$. This reduced purity with respect to the $99.5\%$ purity obtained when all states in a block are, for instance, $\sigma_{z}$ eigenstates, is due to the large voltage differences that are required as an input to the polarization controller for different consecutive states in a block of random states. The limited response time of the involved electronics leads to state generation with non-optimal purity.
Note that, even though we are using an attenuated laser in the experiment, our data also gives us a good estimation of the performance we would obtain with true single photons emitted with an efficiency $\mu$. Indeed (see Methods and Fig.~\ref{fig:losses}), for our regime of parameters, the expected value of the correctness parameter $c$ for single-photon states (blue lines in Fig.~\ref{fig:result2}) is extremely close to those for weak coherent states. Thus, our experimental data can be analyzed for the various input state and quantum memory configurations considered here.

For the ``weak coherent state'' protocol, the security threshold of Eq.~(\ref{eq:cepsilon2}), taking into account USD attacks only, reaches 1 for $\mu \gtrsim 1$. Hence, for larger values of $\mu$, the protocol is insecure against USD attacks. Note that for the threshold of Eq.~(\ref{eq:cepsilonlambda}) which ensures security against any attack (for phase-randomized states), it can be seen that $\mu$ must not exceed $0.40$. The simulations in Fig.~\ref{fig:result2} also show that the protocol is insecure when the state purity $p$ drops below $76\%$, since the value of $c$ then falls below the USD attack security threshold. However, for $0.01 \leq \mu < 1$ and for $p \geq 0.76$, there is a wide region of parameters for which the protocol is secure against such attacks. Indeed, for our experimental results with $p=0.93$, the protocol is secure against USD attacks for values of $\mu$ up to 1.
In Fig.~\ref{fig:losses}, we show the security regions corresponding to Eq.~(\ref{eq:pertes}) as a function of $\mu$ and $\eta_\textrm{tot}=\eta_{\text{qm}}\eta_{\text{det}}$: our experiments, with $\eta_{\text{qm}}=1$ and $\eta_{\text{det}}=0.25$, are situated in the secure region for all values of $\mu \leq 2$.

For the ``single-photon state'' protocol, the security threshold of Eq.~(\ref{eq:cepsilon}) is constant and equal to $7/8=0.875$ for all values of $\mu$. Our experimental data, interpreted as if resulting from single-photon states with a polarization purity $p=0.93$ and an efficiency $0.02 \leq \mu \leq 1$, are secure and show a value of $c$ well above the security threshold. We also notice that the protocol can tolerate large attenuations even for relatively low values of purity.



\begin{figure}
\centering
\includegraphics[width=0.45\textwidth]{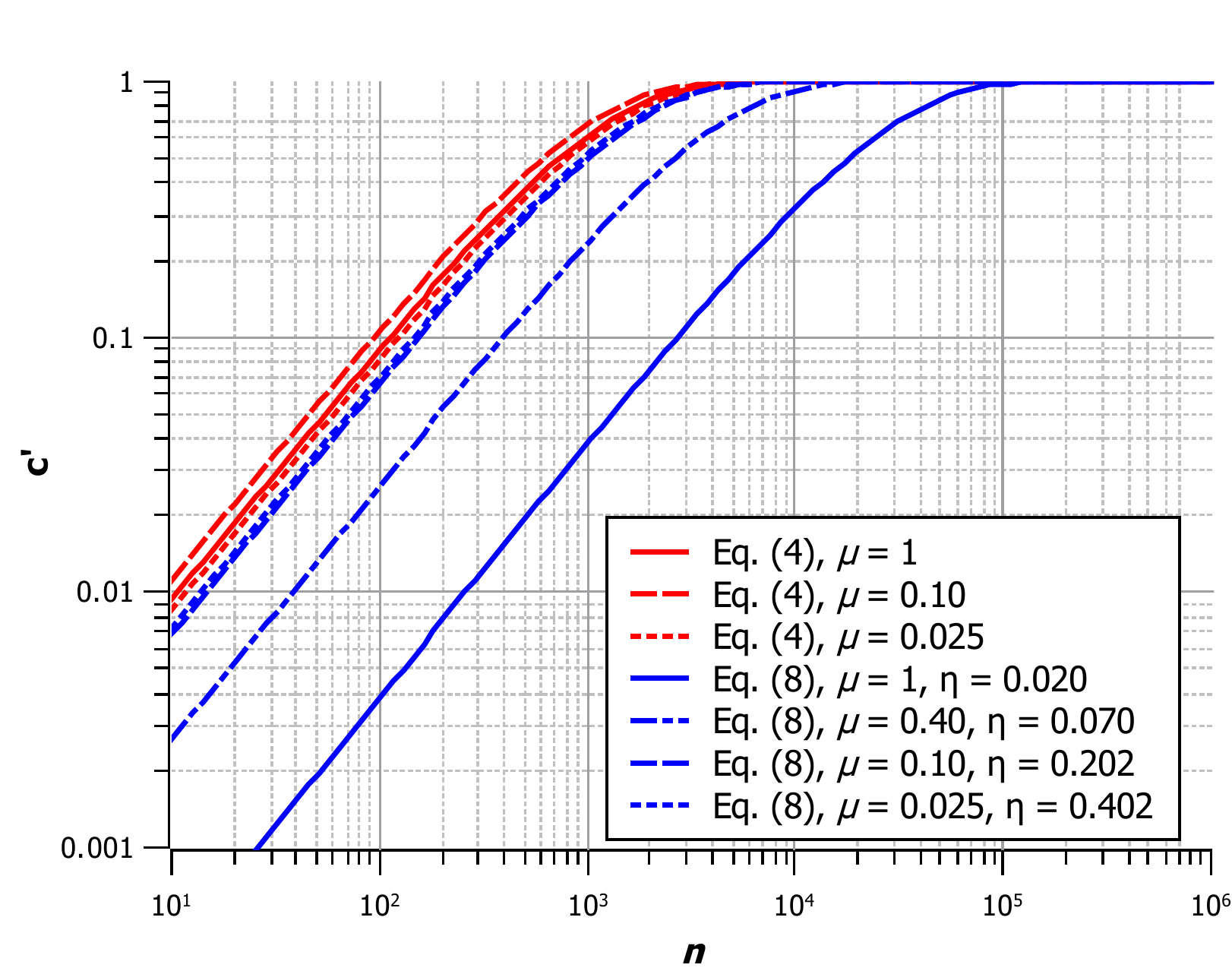}
\includegraphics[width=0.45\textwidth]{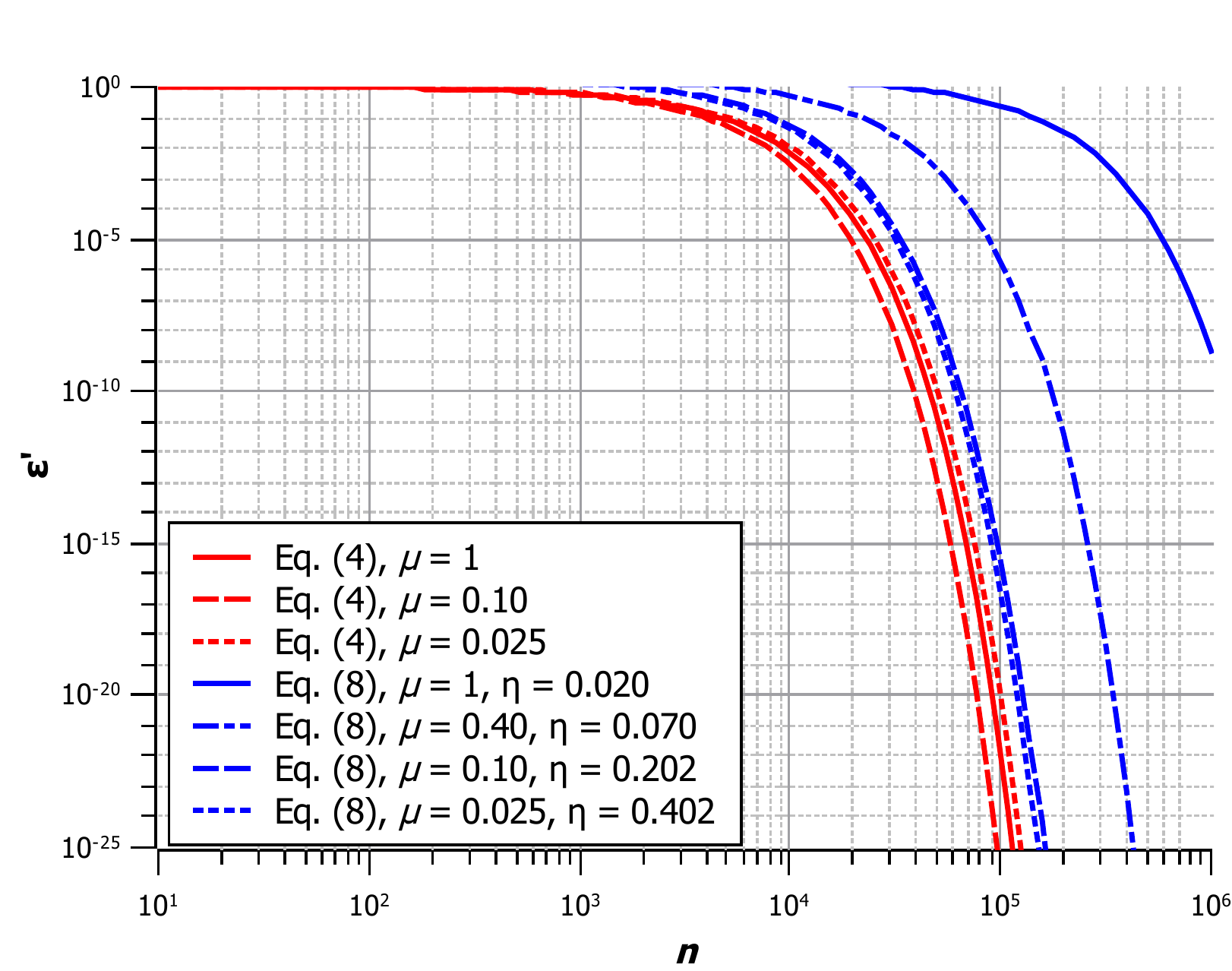}
\caption{\textbf{Correctness and security parameters of the full scheme.} Numerical calculations for the correctness parameter $c'$ (upper graph) and security parameter $\epsilon'$ (lower graph) as a function of the number $n$ of measured qubit pairs in the credit card, with experimental values of $c=0.953 \pm 0.011$ for $\mu=1$ (full red and blue lines), $c=0.965 \pm 0.010$ for $\mu=0.40$ (mixed blue line), $c=0.966 \pm 0.018$ for $\mu=0.10$ (dashed red and blue lines) and $c=0.953 \pm 0.014$ for $\mu=0.025$ (dotted red and blue lines). Note that the lowest values of the $5\sigma$ error bars are considered for plotting these bounds. Red lines correspond to the ``single-photon state'' protocol while blue lines correspond to the ``weak coherent state'' protocol.}
\label{fig:simuvsn}
\end{figure}

\medskip

The measured values for $c$ allow us to estimate the number $n$ of qubit pairs required for our prepaid quantum credit card scheme (corresponding to the game $G'$) to reach a high level of security. In Fig.~\ref{fig:simuvsn}, we show values for the correctness and security parameters $c'$ and $\epsilon'$ defined in Eq.~(\ref{eq:cepsilonprime}) for the ``single-photon state'' protocol, using the experimental values $c=0.953 \pm 0.011$ for $\mu=1$ (full red line), $c=0.966 \pm 0.018$ for $\mu=0.10$ (dashed red line) and $c=0.953 \pm 0.014$ for $\mu=0.025$ (dotted red line). We see that, as $\epsilon'$ drops quickly with the number of qubit pairs $n$, a measured credit card state consisting of a number of pairs comprised between $10^{4}$ and $10^{5}$ is sufficient to reach an arbitrarily small cheating probability in this case, for a wide range of efficiencies $\mu \in [0.025;1]$. Note that when estimating $c'$ and $\epsilon'$, we use the lowest value for the experimental value of $c$ taking into account error bars of $5\sigma$. In this way, there is a probability no higher than $10^{-6}$ that the true $c$ value actually lies beneath this point. 

In Fig.~\ref{fig:simuvsn}, we also display values for the correctness and security parameters $c'$ and $\epsilon'$ defined in Eq.~(\ref{eq:cepsilonprime2}) for the ``weak coherent state'' protocol, using the experimental values $c=0.953 \pm 0.011$ for $\mu=1$ (full blue line), $c=0.965 \pm 0.010$ for $\mu=0.40$ (mixed blue line), $c=0.966 \pm 0.018$ for $\mu=0.10$ (dashed blue line) and $c=0.953 \pm 0.014$ for $\mu=0.025$ (dotted blue line). The parameter $\eta$ has an opposite effect in the two terms in the expression of $\epsilon'$ and we find that these two terms must be roughly balanced. Values for $\eta$ have therefore been chosen accordingly, and we see that the optimal values strongly depend on $\mu$ : they must be increased as $\mu$ decreases. We also notice that, in general, states with large values of $\mu$ require a higher number of detected pairs than states with small values of $\mu$ in order to reach the same security level. However, as long as $\mu$ is not too big, the minimal number of pairs remains of the order of $10^{5}$, and this effect is counter-balanced by the fact that a higher value of $\mu$ increases the number of useful detected pulses and hence the number $n$ of detected pairs. Thus, despite this trade-off, in order to optimize the performance of the setup, it is in general preferable to keep $\mu$ as high as possible in order to maximize the number of detected pairs. We may therefore conclude that our proof-of-principle experiment for the ``weak coherent state'' protocol works optimally when $\mu\in[0.10;0.40]$.\\



\noindent {\large \textbf{Discussion}}\\
\noindent The results that we have presented constitute the first on-the-fly implementation of provably unforgeable quantum money. Our implementation is based on a practical photonic setup with requirements close to those of standard quantum key distribution systems, which is used for quantum credit card on-the-fly generation and readout. The validation of our quantum money protocol and the chosen experimental conditions anticipate the future use of state-of-the-art quantum storage devices, based on single emitters or atomic ensembles, for real-world realization of credit card states.

The integration of our system with a quantum memory requires further developments, in particular to ensure wavelength matching and synchronization, and the full system will be mainly limited by the performance of the quantum storage device, in particular with respect to timing and losses. Regarding losses, we remark that our implementation has minimal channel losses as the transaction is performed locally, while detection losses are processed through our post-selection procedure and taken into account in our security analysis when practical weak coherent states are used in the implementation.

Our work sets a complete theoretical and operational framework for quantum money, and is fully compatible with presently existing quantum memories. In this sense, we provide a crucial experimental benchmark for unforgeability of quantum money, and for any other application where our transaction framework may be relevant.\\


\noindent {\large \textbf{Methods}}\\
\noindent \textbf{Derivation of $\bm \delta$.} For the completeness of game $G'$, we have that the honest client must ensure at least $c-\delta$ of the $Q_{xx}$ challenges, or at least $c-\delta$ of the $Q_{zz}$ challenges, are answered correctly. The correctness $c'$ in the first half of Eq.~(\ref{eq:cepsilonprime}) comes from a simple Chernoff bound, since the client succeeds with probability $c$ in the challenge. In order to prove that the security parameter $\epsilon'$ is the one given in the second part of Eq.~(\ref{eq:cepsilonprime}), it suffices to show that the cheating client has to ensure the challenge $Q_{\epsilon}$ is answered correctly for a fraction of at least $\epsilon+\delta$ of the games $G$ in order to cheat in $G'$. Note that the client must be able to ensure the correct answer for at least a fraction of $c-\delta$ of each of the two challenges and hence the fraction of games where both challenges must be answered is at least $2(c-\delta)-1$, as illustrated below:
\begin{figure}[H]
    \centering
    \includegraphics[width=0.45\textwidth]{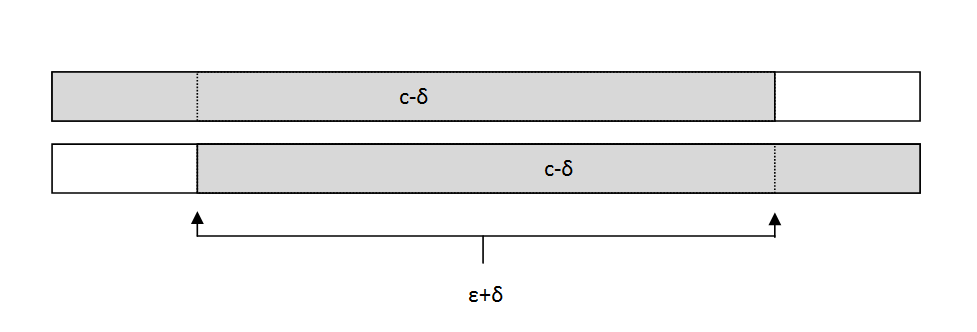}
    \label{fig:simusvsn}
\end{figure}

\noindent Making this equal to $\epsilon+\delta$ provides the value of $\delta$ for the ``single-photon state'' protocol:
\begin{eqnarray}
\epsilon + \delta  & = & 2(c-\delta) - 1\nonumber\\
\delta & = & \frac{2c-\epsilon-1}{3}
\label{eq:delta}
\end{eqnarray}
For the ``weak coherent state'' protocol, taking into account the Poissonian nature of these states, we have that the extra probability $(1+\eta)P_{D}$ of successful USD per pulse goes straight to the adversary. Equation~(\ref{eq:delta}) may then be rewritten as
\begin{equation}
\delta=\frac{2c-\epsilon - (1+\eta)P_{D}-1}{3}.
\end{equation}

\medskip

\noindent \textbf{Setting of $\bm \mu$.} The value of $\mu$ in our experiment is defined as the average photon number per single-photon-detector gate at the output of the polarization controller. It can be expressed as follows:
\begin{equation}
\mu = \frac{\lambda}{hc}\frac{\tau_{\text{gate}}}{\tau_{\text{pulse}}}\frac{\eta_{\text{PC}}\:r_{\text{split}}\:P_{99\text{mean}}}{f_{\text{rep}}},
\end{equation}
where $\lambda=1564$~nm is the wavelength, $h$ is Planck's constant, $c$ is the speed of light in vacuum, $\tau_{\text{gate}}=500$~ps is the duration of the detection gate of the single-photon detectors, $\tau_{\text{pulse}}=20\,\mu$s is the duration of the light pulses, $f_{\text{rep}}=20$~kHz is the repetition rate of these pulses, $r_{\text{split}}$ is the exact splitting ratio of the 99/1 beam splitter, $\eta_{\text{PC}}=0.50$ is the transmission coefficient of the polarization controller and $P_{99\text{mean}}$ is the mean power measured at the 99 output of the 99/1 beam splitter.\\

\medskip

\noindent \textbf{Extraction of $\bm c_{zz}$ and $\bm c_{xx}$.} We estimate the values of the correctness parameters from the experimental results by post-selecting pulses where at least one of the single-photon detectors clicked as follows:
\begin{eqnarray}
c_{zz} &=& \frac{1}{2}(c_0+c_1) \nonumber \\
&=& \frac{1}{2}\frac{N_{\ket{0}}(D_{0/+}\,\text{only})}{N_{\ket{0}}(D_{0/+}\,\text{or}\,D_{1/-})} \nonumber \\
&+& \frac{1}{2}\frac{N_{\ket{1}}(D_{1/-}\,\text{only})}{N_{\ket{1}}(D_{0/+}\,\text{or}\,D_{1/-})},
\end{eqnarray}
and
\begin{eqnarray}
c_{xx} &=& \frac{1}{2}(c_++c_-) \nonumber \\
&=& \frac{1}{2}\frac{N_{\ket{+}}(D_{0/+}\,\text{only})}{N_{\ket{+}}(D_{0/+}\,\text{or}\,D_{1/-})} \nonumber \\
&+& \frac{1}{2}\frac{N_{\ket{-}}(D_{1/-}\,\text{only})}{N_{\ket{-}}(D_{0/+}\,\text{or}\,D_{1/-})},
\end{eqnarray}
where $c_s$ is the fidelity of the state $\ket{s}$, $N_{\ket{s}}(D_{0/+}\,\text{or}\,D_{1/-})$ is the number of pulses corresponding to the state $\ket{s}$ that generated a click on at least one of the detectors, $N_{\ket{s}}(D_{0/+}\,\text{only})$ and $N_{\ket{s}}(D_{1/-}\,\text{only})$ are the number of pulses corresponding to the state $\ket{s}$ that generated a click on the detector $D_{0/+}$ but not on the detector $D_{1/-}$ or on $D_{1/-}$ but not on $D_{0/+}$, respectively, with $s=0,1,+,-$.

The parameter $c_{zz}$ is estimated from measurements performed in the $\sigma_z$ basis for the entire block, \ie, with the half-wave plate rotated to $0^\circ$, while $c_{xx}$ is estimated from measurements performed in the $\sigma_x$ basis for the entire block, \ie, with the half-wave plate rotated to $22.5^\circ$.\\

\medskip

\noindent \textbf{Statistical errors.} The pulse sequences used for measuring $c_{zz}$ and $c_{xx}$ for each value of $\mu < 1$ consisted of $3\times10^6$ pulses (\ie, $1.5\times10^6$ pairs) before post-selection. Each of the four polarization states was generated with probability $0.25$, that is, each state was produced $7.5\times 10^{5}$ times in the random sequence. The total number of post-selected pairs for each block was comprised between $1.3\times 10^{5}$ and $2.6\times 10^{5}$ depending on the value of $\mu$ and for a detection efficiency of $25\%$. Errors on the correctness parameters were estimated by propagating the Poisson errors of the click counting.\\

\medskip

\noindent \textbf{Theoretical models for $\bm c$.} The correctness parameter and its evolution with $\mu$ can be simulated with a simple theoretical model taking into account experimental parameters. This model was used to plot the simulation curves in Fig.~\ref{fig:result2}. We model the polarization state generated by the polarization controller as a density matrix $\rho = p\ket{s}\bra{s} + (1-p)\frac{\Id}{2}$, where $\ket{s}$ is the ideal target state, with $s=0,1,+,-$, $\Id$ is the identity matrix and $p$ is the polarized fraction of the light.

This polarization state is associated with a weak coherent state $\ket{\alpha} =  e^{-\frac{\mu}{2}}\sum_{n=0}^{\infty} \frac{\alpha^n}{\sqrt{n!}}\ket{n}$ with mean photon number $\mu = |\alpha|^2$. For a threshold single-photon detector with detection efficiency $\eta_{\text{det}}$ and dark count probability per detection gate $P_{dc}$, the click probabilities for $\rho$ can be expressed as: $P(D_{s})=P_{dc}+\big(1-e^{-\mu\eta_{\text{det}}}\big)(1+p)/2$ and $P(D_{\bar{s}})=P_{dc}+\big(1-e^{-\mu\eta_{\text{det}}}\big)(1-p)/2$, where $s={0/+}$ and $\bar{s}={1/-}$ if $s=0,+$, and $s={1/-}$ and $\bar{s}={0/+}$ if $s=1,-$. If instead, we consider true single-photon states, with an emission efficiency $\mu$, the click probabilities for $\rho$ are: $P(D_{s})=P_{dc}+\mu\eta_{\text{det}}(1+p)/2$ and $P(D_{\bar{s}})=P_{dc}+\mu\eta_{\text{det}}(1-p)/2$.

In both cases, the state correctness, post-selected on events with at least one click, can be expressed as: $c_s=\frac{P(D_{s})\big(1-P(D_{\bar{s}})\big)}{1-\big(1-P(D_{s})\big)\big(1-P(D_{\bar{s}})\big)}$.

\bigskip


\noindent {\large \textbf{Acknowledgments}}

\noindent We thank Julien Laurat and Frederic Grosshans for useful discussions. This research was supported by the European Research Council project QCC, the French National Research Agency project quBIC and the BPI France project RISQ. 

\medskip










\bibliography{ExpQuantumMoney_bib}

\end{document}